\documentclass[twoside]{dis09}
\usepackage[latin1]{inputenc}
\usepackage[dvips]{graphicx,epsfig,color}
\usepackage{wrapfig,rotating}
\usepackage{amssymb,amsmath,array}

\begin{document}
\title{e+A physics at a future Electron-Ion Collider}

\author{Matthew A. C. Lamont
%
%
\vspace{.3cm}\\
%
Brookhaven National Lab - Physics Department \\
Upton, NY 11973 - USA
}

\maketitle

\begin{abstract}
A future Electron-Ion Colllider (EIC) is the ideal laboratory for studying the gluon distributions in both nucleons and nuclei for $\sqrt{s}$ = 63 - 158 (40 - 110) GeV/A for e+p (e+A) collisions.  Whilst gluon distributions have been studied extensively in nucleons at HERA, there is very little information on them for $x <$ 0.1 in nuclei.  The rapid increase in the gluon momentum distribution at low-$x$ in nucleons found at HERA, if not tamed, leads to the violation of the Froissart Unitarity Bound at small-$x$.  This can be achieved in non-linear QCD by allowing for the recombination of low-$x$ gluons until saturation of gluon densities occurs.  Understanding saturation is an important goal not just for QCD in general, but also in understanding the initial conditions of heavy-ion collisions in particular.  In this paper, I describe the physics that will be explored by e+A collisions, where saturation is more easily explored in this mode at an EIC than in e+p collisions due to the large nuclear densities achieved.
\end{abstract}

Both lattice gauge calculations and effective field theories have shown that the dynamics of the QCD vacuum is dominated by the self-interactions of gluons.  In fact, all of the essential features of QCD are derived from the self-interactions of the gluons.  However, despite their importance in QCD which forms the basis of the standard model, very little is known about the gluons themselves when bound in both nucleons and in nuclei.  In order to understand the role of gluons in nuclei, they can be studied in either e+A or p+A collisions.  However, due to the soft gluon interactions which occur in p+A collisions that modify the nuclear wavefunction and lead to a breakdown of factorization~\cite{Ref:Factorization}, it is more desirable to study e+A collisions and the process of nuclear Deep-Inelastic Scattering (nDIS).

The invariant cross-section in nDIS can be written as:

\begin{eqnarray*}
\frac{d^2 \sigma^{eA \rightarrow eX}}{dx dQ^2} =
\frac{4 \pi \alpha^2_{e.m.}}{xQ^4} \left[ \left(1-y+\frac{y^2}{2} \right )
    F^A_2(x,Q^2) - \frac{y^2}{2} F^A_L(x,Q^2) \right]
\end{eqnarray*}

where $y$ is the fraction of the energy lost by the lepton in the rest frame of the nuclei.  $F^A_2$ represents the quark and anti-quark structure function and $F^A_L$ represents that of the gluons.  Therefore, by measuring the cross-section, it is possible to extract  $F^A_2$ and  $F^A_L$.   $F^A_2$ can be extracted at small $y$ where  $F^A_L$ does not contribute to the cross-section.   $F^A_L$ can then be extracted by running at different values of $y$.

\section{Non-linear QCD and Saturation}

\begin{wrapfigure}{r}{0.5\columnwidth}
\centerline{\includegraphics[width=0.45\columnwidth]{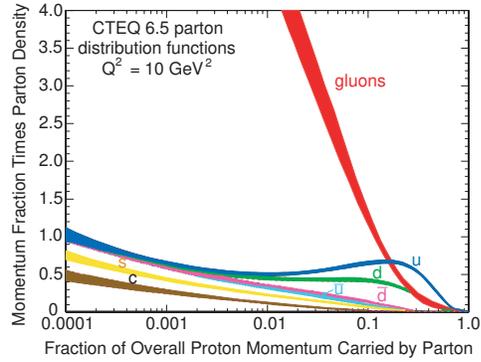}}
\caption{The momentum distributions of quarks and the gluons as a function of x.}\label{Fig:MomDns}
\end{wrapfigure}

The  $F_2$ structure function has been measured in e+p collisions at HERA for a very large range of $x$ and $Q^2$.  By measuring the violations of this scaling and comparing them to NLO QCD fits, it is possible to extract $F_L(x,Q^2) (\propto \alpha_sxG(x,Q^2)$ at low $x)$.  This was performed at HERA for e+p collisions and the corresponding momentum distributions extracted for quarks and gluons as a function of $x$ are presented in Figure \ref{Fig:MomDns}.

This figure is normally shown where the gluon distribution is scaled by a factor of 0.05, but when this is not done, it can be more easily seen that the gluons dominate the distribution for $x$ $<$ 0.1.  At smaller $x$, the gluon distribution rises dramatically, which occurs due to gluon Bremmstrahlung populating the gluons at low $x$ and can be described by linear QCD equations (DGLAP~\cite{Ref:DGLAP} along $Q^2$ and BFKL~\cite{Ref:BFKL} along $x$).  This leads to the conclusion that the Froissart Unitarity Bound will be broken if the growth in the distribution is not tamed.  In processes described by non-linear QCD and the JIMWLK equations~\cite{Ref:JIMWLK}, soft, low-$x$ gluons are allowed to recombine to form higher-$x$ gluons.  When these rates become equal, saturation will occur and this is governed by the saturation scale, $Q^2_{S}$.  The Colour Glass Condensate EFT~\cite{Ref:CGC}, which incorporates saturation, describes particle production at low-$x$ in d+Au collisions at RHIC.

\begin{wrapfigure}{r}{0.5\columnwidth}
\centerline{\includegraphics[width=0.45\columnwidth]{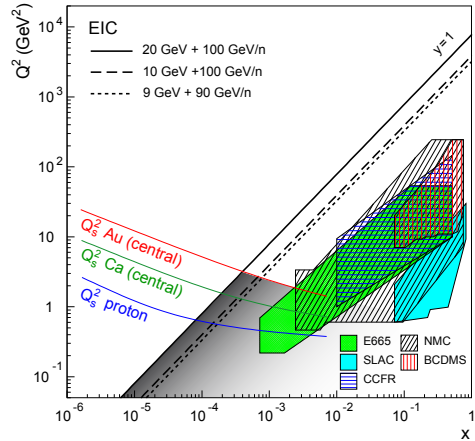}}
\caption{The capabilities of the EIC to reach the saturation regime in e+A collisions.}\label{Fig:QSat}
\end{wrapfigure}

In order to explore this physics, we must study collisions at small-$x$ inside the saturation region.  The capabilities of the proposed EIC, to be built in the US, are shown in Figure \ref{Fig:QSat}, along with the $x$-$Q^2$ phase space which has already been explored in e+A collisions (albeit with very limited statistics).

\subsection{The Nuclear Enhancement Factor}


In order to measure saturation, one has to consider collisions which occur at very small $x$, as shown in Figure \ref{Fig:QSat}.  However, a high-enough $Q^2$ of a few GeV$^2$ is required in order to have reliable theoretical calculations.  Therefore, measuring the saturation regime in e+p collisions at the EIC is ruled out.  However, Figure \ref{Fig:QSat} also shows the saturation scale for e+A collisions and it can be seen that these can be measured at an EIC.  This is because the saturation scale is also determined by geometry at leading order, where $Q^2_S$ $\propto$ $(A/x)^{1/3}$ and therefore one can perform e+A collisions for heavy A which will reach a much lower effective x than in e+p collisions at the same energy.  The increased capabilities that this offers means that e+A collisions are the ideal laboratory for measuring saturation physics in an experimental regime.

\section{The four key measurements for the e+A physics programme}

The key measurements of the e+A physics programme are as follows:

\subsection{What is the momentum distribution of gluons in nuclei?}

The $F^A_2$ structure function is determined by the momentum distributions of the quarks and anti-quarks.  By measuring the scaling violation of $F^A_2$ with $Q^2$, $F^A_L$ can be determined which, in QCD, is directly proportional to the gluon structure ($F^A_L \propto  \alpha_S xG(x,Q^2)$.  This has been extensively studied at HERA with respect to protons, but is unknown for nuclei.

The measurement of $F^A_L$ though the scaling violation of $F^A_2$ is an indirect measurement.  However, it is possible to measure $F^A_L$ directly.  In order to do this, it is required that measurements are taken at different beam energies and hence different $y$.  This has also been performed with small statistics at HERA in the case of protons but was only possible due to the different energies run at the end of HERA's lifetime.

\begin{wrapfigure}{r}{0.5\columnwidth}
\centerline{\includegraphics[width=0.45\columnwidth]{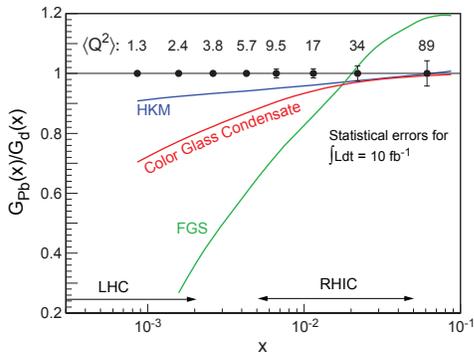}}
\caption{The $G_{Pb}(x)/G_{d}(x)$ ratio measured at the EIC as a function of $x$.}\label{Fig:F_L}
\end{wrapfigure}

At the EIC, it will be possible to run both the lepton and hadron beams in various configurations.  Figure \ref{Fig:F_L} shows the ratio of G$(x)$ measured in $e$+Pb collisions to that in $e$+$d$ collisions.  The predicted statistical error bars achievable at an EIC for a 2 year running period (total $\int L$ = 10/A $fb^{-1}$), 4/A $fb^{-1}$ at 10+100 GeV/A, 4/A $fb^{-1}$ at 10+50 GeV/A and 2/A $fb^{-1}$ at 5+50 GeV/A are shown on the plot.  Three different models are shown on the plot where it is evident that the statistical error bars obtained from a direct measurement of $F^A_L$ will allow for a differentiation between models, particularly at low-$x$.  However, the systematic errors on this measurement are expected to be significant and are yet to be estimated.

\subsection{What is the space-time distribution of gluons in nuclei?}

Not only do we wish to know the momentum distribution of the gluons, we also wish to know the spatial distribution of gluons - are they distributed uniformly or are they distributed in clumps?  The spatial distribution of gluons can be extracted by viewing the collision in the frame in which the virtual photon fluctuates into a quark-anti-quark dipole, which then subsequently scatters coherently on the nucleus.

The optical theorem is used to calculate the survival probability, that is, the probability of the dipole to travel through the nucleus without interacting.  In pQCD, this survival probability is close to unity.  This is in contrast to dipole models which have this as low at 20$\%$,  depending on the impact parameter.


\subsection{What is the role of colour-neutral excitations?}

At HERA, in e+p collisions, the surprising result was found that in $10-20\%$ of the collisions, the struck nucleon remains intact in the final state.  These events are referred to as diffractive and occur when the electron interacts with a colour neutral vacuum excitation, which in QCD can be seen as a combination of two or more gluons, called the pomeron.  In e+A collisions at the EIC, this is predicted to occur up to 40$\%$ of the time~\cite{Ref:Navarro}.  The measurement of exclusive vector meson production in diffractive events is important as their cross-section $\propto \alpha_sxG(x,Q^2)^2$, that is, even more sensitive to the gluon distribution than $F^A_L$.  

However, despite it's sensitivity, the measurement of diffractive collisions are an experimental challenge.  For the case of coherent diffraction at small $t$ ($\approx <$ 300 MeV$^2$~\cite{Ref:STARUPC}), the struck nucleus will remain intact.  For larger values of $t$, the nucleus will break-up, though it's quantum numbers will remain in the final state.  In order to measure the nucleus, it must be measured in a detector.  Currently, the best method for this is with ``Roman Pot" detectors which are placed downstream of the interaction point and which go close to the beam.  However, there are limitations on how close they can go to the beam, which is generally considered to be approximately 10 times the angular divergence of the beam, where $\theta_{beam} = \sqrt{\epsilon/\beta^*}$, where $\epsilon$ is the beam emittance and $\beta$ the amplitude/betatron function.

\begin{wrapfigure}{r}{0.5\columnwidth}
\centerline{\includegraphics[width=0.45\columnwidth]{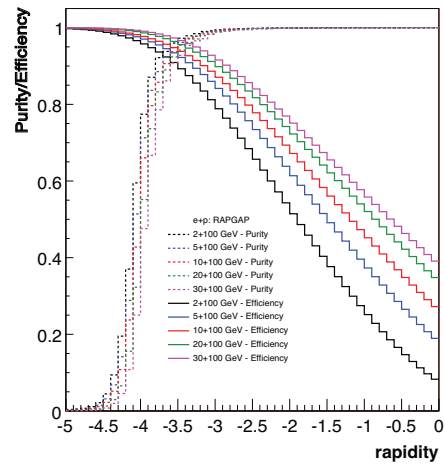}}
\caption{The purity and efficiency of measuring diffractive events for an ideal detector.}\label{Fig:PurityEfficiency}
\end{wrapfigure}

In order to achieve scattering angles larger than the angular divergence, the nucleus needs to survive a $p_T$-kick greater than $p_T^{min} \approx pA\theta_{min}$.  For a beam energy of 100 GeV/n, $\beta^*$ = 60 m and $\theta_{min}$ = 0.08 mrad, then $p_T^{min}$ for Au is 1.58 GeV/c.  This is much greater than the energy required to break up the nucleus and therefore, the intact nucleus cannot be detected.  This will be possible for lighter nuclei, however, where, for Si,  $p_T^{min}$ $\approx$ 0.22 GeV/c.

It is still possible to measure diffractive events though through the existence of rapidity gaps between the outgoing nuclei and the most forward hadron in the event.  These rapidity gaps were used at HERA to distinguish diffractive events from normal DIS events.  Diffractive and DIS events were simulated at possible EIC energies using the RAPGAP MC generator for e+p collisions and a study performed using the large rapidity gap method to distinguish event classes.  Figure \ref{Fig:PurityEfficiency} shows the efficiency (number of diffractive events out of all diffractive events which pass cut) and purity (number of diffractive events out of all events which pass cut) of the diffractive event sample when assuming that the cross-section is 1/3 of the total cross-section.  This is almost independent of collision energy and for all cases, high efficiencies and high purities can be obtained simultaneously.  

\subsection{What are the mechanisms of hadronization and energy loss in cold nuclear matter?}

\begin{wrapfigure}{r}{0.5\columnwidth}
\centerline{\includegraphics[width=0.45\columnwidth]{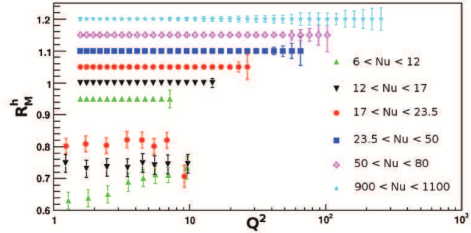}}
\caption{Hadronic energy loss capabilities in the EIC and HERMES.}\label{Fig:Had}
\end{wrapfigure}

One of the important results emanating from RHIC is the suppression of particles with high transverse momentum and the suppression of particles in jets in A+A collisions~\cite{Ref:JetQuenchingSTAR}.  This is quite striking, a factor of 5 compared to what would be expected from scaled p+p collisions.  Initially this was explained by energy loss through the radiation of gluons in the hot and dense partonic medium.  However, heavier charmed hadrons also show the same amount of suppression which cannot be explained through this process as the radiation of gluons is suppressed at forward angles for heavier particles, a process known as the ``Dead-Cone Effect".

Therefore, in order to fully understand and interpret the data from RHIC, measurements must be made of the amount of energy loss in cold nuclear matter through e+A collisions.  Measurements were performed by HERMES on light nuclei~\cite{Ref:HERMES}, but the statistics were too low to distinguish between different time-scales on the energy loss.  Figure \ref{Fig:Had} shows the capabilities of the EIC plotted together with what was possible at HERMES.  The EIC capabilities far out-weigh those of HERMES in terms of both statistics, $Q^2$ reach and $\nu$.

\section{Summary and Conclusions}

In summary, the role of gluons is one of the most important, yet least understood contributions to QCD.  In order to understand the surprising results to come out of HERA, it is imperative that further studies are made at an Electron-Ion Collider.  Such a collider must have the capabilities to accelerate heavy-ions in order to give access to the effective-$x$ where saturation is important.   This will help in understanding fundamental QCD physics as well as understanding the initial conditions of heavy-ion collisions.  Such a proposal, put forward in the US, would be able to address the questions posed in this manuscript.

\begin{footnotesize}




\begin{thebibliography}{99}
\bibitem{Ref:Factorization}F.-P. Schilling, Acta Phys. Polon. \textbf{B 33} (2002) 3419, hep-ex/0209001.
\bibitem{Ref:DGLAP}V.~N. Gribov and L.~N. Lipatov,
\newblock Sov. J. Nucl. Phys. {\bf 15}, 438 (1972). \\
G.~Altarelli and G.~Parisi,
\newblock Nucl. Phys. {\bf B126}, 298 (1977). \\
Y.~L. Dokshitzer,
\newblock Sov. Phys. JETP {\bf 46}, 641 (1977).
\bibitem{Ref:BFKL}E.~A. Kuraev, L.~N. Lipatov, and V.~S. Fadin,
\newblock Sov. Phys. JETP {\bf 45}, 199 (1977). \\
I.~I. Balitsky and L.~N. Lipatov,
\newblock Sov. J. Nucl. Phys. {\bf 28}, 822 (1978).
\bibitem{Ref:JIMWLK}J.~Jalilian-Marian \textit{et al.}, Phys. Rev. {\bf D59}, (1999) 014014.
\bibitem{Ref:CGC}E.~Iancu and R.~Venugopalan, review for QGP3, Eds. R. C. Hwa and X.-N. Wang, World Scientific (2003), arXiv:hep-ph/0303204.
\bibitem{Ref:Navarro}M. S. Kugeratski, V. P. Goncalves, F. S. Navarra, Eur. Phys. J. {\bf C 46} (2006) 465.
\bibitem{Ref:STARUPC}B. Abelev  $et ~al.$ [STAR Collaboration], Phys. Rev. {\bf C 77} (2008) 34910.
\bibitem{Ref:JetQuenchingSTAR}C. ~Adler $et al.$ [STAR Collaboration], Phys. Rev. Lett. \textbf{89} (2002) 202301.
\bibitem{Ref:HERMES}A. Airapetian {\it et~al.} [HERMES Collaboration] Phys. Lett. {\bf B 577} (2003) 37.
\end{thebibliography}
%

\end{footnotesize}


\end{document}